\documentstyle[12pt,aasms]{article}
\tighten

\slugcomment{Accepted to MNRAS}
\begin{document}

\title{Paramagnetic alignment of thermally rotating dust}
\author{A. Lazarian  }
\altaffiltext{1}{Department of Astrophysical Sciences, Princeton University,
Princeton, NJ 08544}

\begin{abstract}
Paramagnetic alignment of  
thermally rotating oblate dust grains 
is studied analytically 
for finite ratios of grain to gas temperatures.  
For such ratios, the alignment of angular momentum $\bf J$ 
in respect to the
 grain axis of maximal
 inertia is only partial.
 We treat the alignment of $\bf J$ using  perturbative methods 
 and disentangle the problem of $\bf J$ alignment
in grain body axes from that of $\bf J$ alignment in respect to 
magnetic field. This enables us to find the alignment of grain
axes to magnetic field and 
thus relate our theory to polarimetric
observations.
Our present results  are 
applicable to the alignment of both paramagnetic and superparamagnetic
grains.

\end{abstract}


\section{Introduction}

Grain alignment in molecular clouds remains a puzzle in spite of
intensive experimental research in the area
(see Whittet 1992, Goodman et al. 1995, Dotson 1996, 
Hildebrand \& Dragovan 1995). We believe,
that one reason for this is that the theory of the alignment 
is still not adequate (see a recent review by Roberge 1996).

Paramagnetic alignment of thermally rotating grains\footnote{A term {\it
thermally
rotating} means that a grain is rotating at Brownian velocities. This is 
possible whenever
grains are not subjected to uncompensated torques (Purcell 1979,
Draine \& Weingartner 1996a,b).}  discovered by
Davis-Greenstein in 1951 is often named a candidate for
explaining  the alignment in molecular clouds. Since then numerous studies
of the alignment were performed (e.g. 
 Jones \& Spitzer 1967, Purcell 1969, 
Purcell \& Spitzer 1971). However the important effect of internal
relaxation was described by Purcell only in 1979. He found that the
angular momentum $\bf J$ is aligned with the grain axis 
of maximal moment of inertia,  ${\bf z}^b$, (henceforth the axis of 
major inertia)
on a time scale much shorter than the gaseous damping time. Both 
the recent numerical (Roberge, DeGraff \& Flaherty 1993) and
analytical (Lazarian 1995a, further on Paper~I) studies of the alignment
of thermally rotating oblate grains\footnote{The 
correspondence between the analytical and numerical results was established 
in DeGraff, Roberge \& Flaherty (1997).} accounted for thermal relaxation
by assuming $\bf J$ and ${\bf z}^b$ to be parallel. This assumption,
however, is valid only when grain rotational temperature is much greater
than its material temperature. 

The fact that for finite ratios of the said temperatures the internal
alignment is not perfect was pointed out in Lazarian (1994), while the 
quantitative study of this effect was given in Lazarian \& Roberge
(1997) (henceforth LR).
The purpose of the present paper is to incorporate the effect of incomplete
internal relaxation into the theory of paramagnetic alignment of oblate 
grains by using perturbative approach.

In Section 2, we formulate the problem. In Sections
3 and 4 we determine the Fokker-Planck coefficients arising from grain-gas
interactions, while magnetic coefficients and the Fokker-Planck
equation for the Barnett relaxation are reproduced in Section 5. The
iterations to the measure of grain axis alignment are obtained in
Section 6, and the summary of main results is presented in Section 7.

\section{Formulation of the problem}

Starlight polarization is caused by the alignment of grain axes, while 
dynamical evolution
is defined in terms of angular momentum. The importance of relating these
 two different quantities was realized by researchers very early
(see Davis \& Greenstein 1951, Davis 1955,
 Jones \& Spitzer 1967, Purcell \&
Spitzer 1971). In the present paper we relate these quantities
when  $\bf J$ is partially aligned with the grain axis of major inertia.

For an ensemble of spheroidal grains the measure of axis alignment in 
respect to magnetic field can be described by the Rayleigh reduction
factor (Greenberg 1968)
\begin{equation}
R=\frac{3}{2}\langle \cos^{2}\phi -\frac{1}{3} \rangle~~~,
\label{one}
\end{equation}
where $\phi$ is the angle between the axis of an oblate grain and 
magnetic field $\bf H$ and, here and further on, angular brackets 
$\langle \rangle$ denote the ensemble averaging.
Similarly, the alignment of angular momentum in grain body axes is given by 
\begin{equation}
Q_{\rm X}=\frac{3}{2}\langle \cos^{2}\theta -\frac{1}{3} \rangle~~~,
\label{two}
\end{equation}
where $\theta$ is the angle between the axis of major inertia and $\bf J$. 
The alignment of $\bf J$ in respect to magnetic field is characterized by
\begin{equation}
Q_{\rm J}=\frac{3}{2}\langle \cos^{2}\beta -\frac{1}{3} \rangle~~~,
\label{three}
\end{equation}
where $\beta$ is the angle between $\bf J$ and $\bf H$.  
These three measures are not independent.
Indeed, it is obvious from spherical trigonometry 
(see eq. 108,  Davis \& Greenstein 1951), that 
\begin{equation}
\langle \cos^{2}\phi\rangle=\frac{1}{2}\left (1-\langle \cos^{2}\beta
\rangle-\langle \cos^{2}
\theta\rangle +3\langle \cos^{2}\beta \cos^{2}\theta\rangle\right )~~~,
\label{trig}
\end{equation}
and we use this identity to relate $R$, $Q_{\rm X}$ and $Q_{\rm J}$. 

In the zeroth order
 approximation the alignment in grain axes is independent of the
alignment
in respect to magnetic field\footnote{Further in the text we discuss
 the dependence between the distributions of $\beta$ and $\theta$ due to their 
dependence on the distribution of $J$.}. Therefore 
\begin{equation}
\langle \cos^{2}\beta \cos^{2}\theta\rangle \approx \langle \cos^{2}\beta
\rangle
\langle \cos^{2}\theta\rangle~~~,
\end{equation}
and it is easy to see that (Jones \& Spitzer 1967)
\begin{equation}
R\approx Q_{\rm J}\times Q_{\rm X}~~~.
\label{rail}
\end{equation}

Here, as in Paper I, we consider only oblate  grains, because
there are indications that aligned grains are oblate rather than prolate 
(Aitken et al. 1985, Lee \& Draine 1985, Hildebrand 1988, Hildebrand 
\& Dragovan 1995). We approximate the grain mantle surface and 
the core-matle interface by confocal spheroids and use 
$a_{m}$ and $b_{m}$ ($b_{m}>a_{m}$) to denote the mantle semi-axes  
parallel and perpendicular to the grain symmetry axis, respectively. 
The corresponding core semi-axes are denoted by 
$a_{c}$ and $b_{c}$. Then the eccentricity of the core/mantle ($i=c,m$) is 
\begin{equation}
{\rm e}_{i}=\sqrt{1-\frac{a_{i}^{2}}{b_{i}^{2}}}~~~.
\end{equation}
It may be different for different components of the grain.

\section{Gaseous bombardment}

It was shown in Jones \& Spitzer (1967) that alignment of $\bf J$ due to 
 paramagnetic relaxation can be 
described by the Fokker-Planck equation (see Reichl 1980)
\begin{equation}
\frac{\partial f}{\partial t}+\frac{\partial}{\partial J_{i}}
\left(\left\langle
\frac{\bigtriangleup J_{i}}{\bigtriangleup t}\right\rangle f\right)=
\frac{1}{2}
\frac{\partial^{2}}{\partial J_{i}\partial J_{j}}
\left(\left\langle
\frac{\bigtriangleup J_{i}\bigtriangleup J_{j}}{\bigtriangleup t}
\right\rangle f\right)~~~,
\label{d32}
\end{equation}
where $f$ is the distribution function of angular momentum $J\equiv 
|{\bf J}|$, while
$\left\langle\frac{\bigtriangleup J_{i}}{\triangle t}\right\rangle$ and 
$\left\langle\frac{\bigtriangleup J_{i}\bigtriangleup J_{j}}{\triangle t}\right\rangle$ 
are diffusion coefficients.

The quadratic diffusion coefficients in the grain frame of reference
$x^{\rm b}y^{\rm b}z^{\rm b}$ were calculated in Roberge et al. (1993):
\begin{equation}
\left\langle\frac{(\triangle J_{z}^b)^2}{\triangle t}\right\rangle=
\frac{2\sqrt{\pi}}{3}nmb_{m}^{4}v_{\it th}^{3}\Gamma_{\|}({\rm e}_m)
\left(1+\frac{T_{\rm s}}{T_{g}}\right)~~~,
\label{3.1}
\end{equation}
\begin{equation}
\left\langle\frac{(\triangle J_{i}^b)^2}{\triangle t}\right\rangle=
\frac{2\sqrt{\pi}}{3}nmb_{m}^{4}v_{\it th}^{3}\Gamma_{\bot}({\rm e}_m)
\left(1+\frac{T_{\rm s}}{T_{g}}\right)~~~,
\label{3.2}
\end{equation}
where $i=x,y$, $T_{\rm s}$ and $T_{\rm g}$ are dust and gas temperatures, respectively,
and $v_{\rm th}=\sqrt{2kT_{\rm g}/m}$ is the thermal velocity of gaseous 
atoms with mass $m$ and concentration $n$. 
The coefficients $\Gamma_{\perp}({\rm e}_{m})$ and 
$\Gamma_{\|}({\rm e}_{m})$ are geometrical factors
\begin{equation}
\Gamma_{\bot}({\rm e}_{m})=
\frac{3}{32}\{7-{\rm e}_{m}^{2}+
(1-{\rm e}_{m}^{2})g_{m}({\rm e}_{m})+(1-2{\rm e}_{m}^{2})
[1+{\rm e}_{m}^{-2}(1-[1-{\rm e}_{m}]^{2}g({\rm e}_{m}))]\},
\label{3.4}
\end{equation}
\begin{equation}
\Gamma_{\|}({\rm e}_{m})=
\frac{3}{16}\{3+4(1-{\rm e}_{m}^{2})g_{m}({\rm e}_{m})-{\rm e}_{m}^{2}
(1-[1-{\rm e}_{m}^{2}]^{2}g({\rm e}_{m}))\},
\label{3.5}
\end{equation}
with
\begin{equation}
g({\rm e}_{m})=\frac{1}{2{\rm e}_{m}}\ln
\left(\frac{1+{\rm e}_{m}}{1-{\rm e}_{m}}\right).
\label{3.6}
\end{equation}
We cannot use the expressions for 
$\left\langle \frac{\triangle J_i}{\triangle t}\right\rangle$ 
obtained in Roberge et al. (1993) as those are found assuming perfect Barnett
alignment. Instead we will derive these coefficients from 
$\left\langle(\frac{\bigtriangleup J_{i i}}{\triangle t})^{2}\right\rangle$
using the approach from Lifshitz \& Pitaevskii (1981, Chapter II).
In thermal equilibrium, the distribution of angular momentum in grain
frame of reference is given by 
\begin{equation}
f={\rm const}\times \exp\left(-\frac{J_x^2+J_y^2}{2I_{\bot}kT_g}-
\frac{J_z^2}{2I_{z}kT_g}\right )
\label{3.7}
\end{equation}
and is independent of grain magnetic properties. 
Then, the terms in the Fokker-Plank equation (\ref{d32})
corresponding to the Davis-Greenstein relaxation, Barnett relaxation and
to the gaseous bombardment can be studied separately. Indeed, these are
independent processes and in thermodynamic equilibrium 
fluctuations of each of these
parameters should be compensated by the corresponding dissipation.
In physical terms this means that neither 
variations of ambient gaseous pressure, nor variations of 
grain magnetic properties change the equilibrium distribution (\ref{3.7}).   
Therefore by plugging Eq.~(\ref{3.7}) in Eq.~(\ref{d32}), we obtain  
\begin{equation}
-\left\langle \frac{\triangle J_i^b}{\triangle t}\right\rangle=
\frac{1}{2I_ikT}
\left\langle\frac{(\triangle J_{i i}^b)^2}{\triangle t}\right\rangle J_i~~~.
\label{3.8}
\end{equation}
Substituting $T_{\rm g}=T_{\rm s}$ in Eqs~(\ref{3.1}) and (\ref{3.2}), 
we get\footnote{Note that 
$\left\langle \frac{\triangle J_i^b}{\triangle t}\right\rangle$ 
is independ of grain temperature. Therefore Eqs~(\ref{3.9}) present 
a general form of  $\left\langle 
\frac{\triangle J_i^b}{\triangle t}\right\rangle$
valid for any $T_{\rm g}$ and $T_{\rm d}$. Another way of deriving
Eqs~(\ref{3.9}) is presented in LR.}
\begin{eqnarray}
\left\langle \frac{\triangle J_z^b}{\triangle t}\right\rangle
=-\frac{4\sqrt{\pi}}{3I_z}nmb_{m}^{4}v_{\it th}\Gamma_{\|}({\rm e}_m)J_z^b ~~~,
\nonumber\\ 
\left\langle \frac{\triangle J_j^b}{\triangle t}\right\rangle
=-\frac{4\sqrt{\pi}}
{3I_{\bot}}nmb_{m}^{4}v_{\it th}\Gamma_{\bot}({\rm e}_m)J_j^b~~~,
\label{3.9}
\end{eqnarray}
 where $j=x, y$. 
In the limit of perfect Barnett relaxation $J_x^b= J_y^b= 0$,  
and Eqs~(\ref{3.9}) reduce to the expressions for the 
diffusion coefficients obtained in Roberge et al. (1993).
Introducing time-scales 
\begin{eqnarray}
t_{gas1}&=&\frac{3I_z}{4\sqrt{\pi}}\frac{1}{nmb_{m}^{4}v_{\it th}
\Gamma_{\|}({\rm e}_m)}~~~, \nonumber\\
t_{gas2}&=&\frac{3I_z}{4\sqrt{\pi}}\frac{1}{nmb_{m}^{4}v_{\it th}
\Gamma_{\bot}({\rm e}_m)}~~~,
\label{3.10}
\end{eqnarray}
it is possible to rewrite expressions (\ref{3.9}) as follows
\begin{eqnarray}
\left\langle \frac{\triangle J_z^b}{\triangle t}\right\rangle
 &=&-J_z^b/t_{gas1} ~~~,
\nonumber\\ 
\left\langle \frac{\triangle J_j^b}{\triangle t}\right\rangle
 &=&-J_j^b/t_{gas2}~~~, 
\label{3.11}
\end{eqnarray}
where $j=x,y$.

The component of the moment of major inertia $I_z$ can be as much as two times
greater than $I_{\bot}$ for extremely flat grains. At the same time
$\Gamma_{\|}$ is only slightly greater than $\Gamma_{\bot}$. Therefore
for any eccentricity,  $t_{gas1}>t_{gas2}$ (see Fig.~2).
This corresponds to an intuitive perception that an oblate grain
rotating about its axis of major inertia experiences less friction
than the one rotating about a perpendicular axis. Indeed, the friction
torque is proportional to grain's angular velocity $\sim J/I_i$, where
$i=x,y$ or $z$ depending on the axis of rotation. Thus for a fixed
$J$ the drag is inversely proportional to the moment of inertia
in accordance with Eq.~(\ref{3.10}).

In general, due to the difference between $t_{gas1}$ and $t_{gas2}$ 
the vector of the dissipation angular momentum is not directed
along $\bf J$. Instead it lies in the plane defined by $\bf J$ and 
axis of major inertia. If grain rotational temperature is determined
by gas-grain collisions, the difference in ``parallel'' and ``perpendicular''
damping is compensated by the difference in excitation of rotation
through the quadratic terms. However, for suprathermal rotation,
this difference ($t_{gas1}>t_{gas2}\;$!)
 should provide alignment of $\bf J$ with the axis of major inertia.
 For paramagnetic grains, this mechanism
is less efficient as compared with the Barnett relaxation (see Section~6.2), 
but there may be situations, where it is important. A further discussion of 
this interesting phenomenon will be given elsewhere.

\section{Averaging over precession}

Grains perform complex motion: for one thing, the grain axis of major
inertia  ${\bf z}^b$ precesses about $\bf J$, 
for another thing, $\bf J$ precesses about magnetic field. 
The period of the latter is much greater ($\sim 10^{10}$ times)
 than the period of the former precession, but is small compared to the 
gaseous damping time. Therefore, following usual approach
adopted in theoretical mechanics to ``fast'' tops (see Landau
\&  Lifshitz 1976), we shall at first 
neglect ambient magnetic field\footnote{We neglect the magnetic field
while averaging, but account for its action in terms of 
decreasing the value of angular momentum.} and consider averaging over free precession
of ${\bf z}^b$ about the direction of the angular
momentum $\bf J$. Then the averaging over Larmor precession (i.e.
precession of $\bf J$ about magnetic field) will be
performed.

\subsection{Averaging over precession about the axis of major inertia}

The angular momentum rapidly precesses in grain body axes and the angular 
velocity of such precession is of the order of grain angular velocity.

It is convenient to introduce a new Cartesian reference
system x,y,z, where $\bf z$ is directed along $\bf J$ and $\bf x$ and
$\bf y$ lie in the plane normal to $\bf J$. The relative orientation 
of xyz and the body frame is given by Eulerian angles (see Fig.~1). 
For an oblate spheroid, $x^b$-axis 
can be taken along the lines of nodes, i.e. $\psi=0$, which greatly simplifies
further calculations. 

The increments of the angular momentum along $x^b$, $y^b$, and $z^b$ axes
are related to the increments along $x$, $y$, and $z$ axes in the
following way:
\begin{eqnarray}
\triangle J_x &=&
\triangle J^{b}_x \cos\phi - \triangle J^{b}_y \cos\theta \sin\phi
+\triangle J^{b}_z \sin\theta \sin\phi~~~, \nonumber\\
\triangle J_y &=&
\triangle J^{b}_x\sin\phi+\triangle J^{b}_y \cos\theta\cos\phi
-\triangle J^{b}_z \sin\theta\sin\phi ~~~, \\
\triangle J_z &=&
\triangle J^{b}_y \sin \beta + \triangle J^{b}_z \cos\beta~~~.
\nonumber
\label{a.1}
\end{eqnarray}

This approach can provide the ``precession avegared'' coefficients
for arbitrary grains. However, to simplify the resulting formulae we
account for the grain rotational 
symmetry and introduce the following notation for the
coefficients in the body axes
\begin{eqnarray}
\left\langle\frac{(\triangle J_{\bot})^2}{\triangle t}\right\rangle &=&
\left\langle\frac{(\triangle J_{x}^b)^2}{\triangle t}\right\rangle = 
\left\langle\frac{(\triangle J_{y}^b)^2}{\triangle t}\right\rangle ~~~,
 \nonumber\\
\left\langle\frac{(\triangle J_{\|})^2}{\triangle t}\right\rangle &=&
\left\langle\frac{(\triangle J_{z}^b)^2}{\triangle t}\right\rangle~~~.
\label{a.1.b}
\end{eqnarray}
Averaging over $\phi$ gives
\begin{eqnarray}
\left\langle \frac{\triangle J_i}{\triangle t}\right\rangle &=&0~~~,
 \nonumber\\
\left\langle \frac{\triangle J_z}{\triangle t}\right\rangle &=&
\left\langle \frac{\triangle J_z^b}{\triangle t}\right\rangle\cos\theta+
\left\langle \frac{\triangle J_y^b}{\triangle t}
\right\rangle \sin\theta~~~,
\label{a.2}
\end{eqnarray}
where $i=x,y$. Using Eq.~(\ref{3.11}), 
it is possible to rewrite Eq.~(\ref{a.2}) as
\begin{eqnarray}
\left\langle \frac{\triangle J_i}{\triangle t}\right\rangle &=&0~~~,
 \nonumber\\
\left\langle \frac{\triangle J_z}{\triangle t}\right\rangle &=&
-\frac{J_z^b}{t_{gas1}}
\cos\theta-\frac{J_y^b}{t_{gas2}}\sin\theta~~~,
\label{a.2.b}
\end{eqnarray}
which shows, that the averaged vector of the gaseous damping is 
directed along $\bf J$\footnote{The component of damping force 
orthogonal to $\bf J$ produces insignificant nutations of $\bf J$. This 
is in contrast to its action in the grain frame of reference, 
where it can contribute
to the alignment of $\bf J$ with the axis of major inertia.}.
 As we have chosen $x^b$-axis  along the 
lines of nodes, the angular momentum is 
\begin{equation}
{\bf J}=J^b_y {\bf i}_y^b+J^b_z {\bf i}_z^b~~~,
\label{a.2.c}
\end{equation}
where $ {\bf i}_y^b$ and $ {\bf i}_z^b$ are unit vectors along 
$y^b$ and $z^b$-axes, respectively, in the grain frame of reference.
It is easy to see, that
\begin{equation}
\tan\theta=\frac{J^b_y}{J^b_z}~~~.
\label{a.2.d}
\end{equation}
Combining Eqs~(\ref{a.2.b}), (\ref{a.2.c}) and  (\ref{a.2.d}),  
it is possible to introduce the effective damping time
\begin{equation}
t_{\rm eff}=t_{gas1}\cdot \xi(\theta)~~~,
\label{a.2.e}
\end{equation}
where
\begin{equation}
\xi(\theta)=\frac{1+\tan^2\theta}{1+\frac{t_{gas1}}{t_{gas2}} \tan^2\theta}~~~.
\label{a.2.eee}
\end{equation}
According to Fig.~2, $\frac{t_{gas1}}{t_{gas2}}>1$, and therefore
$t_{\rm eff}<t_{gas1}$.
However for grains of small eccentricities and/or  $\theta\ll 1$, 
the difference between $t_{\rm eff}$ and $t_{gas1}$ is small. Using 
$t_{\rm eff}$ it is possible to rewrite Eq.~(\ref{a.2.b}) in the following way
\begin{equation}
\left\langle \frac{\triangle J_z}{\triangle t}\right\rangle =
-\frac{J_z}{t_{\rm eff}}~~~.
\label{a.2.e1}
\end{equation}

After simple algebra we obtain the ``precession averaged''
diagonal quadratic coefficients of the Fokker-Planck equation:
\begin{eqnarray}
\left\langle\frac{(\triangle J_{i})^2}{\triangle t}\right\rangle &=&
\frac{1}{2}\left (
\left\langle\frac{(\triangle J_{\bot})^2}
                 {\triangle t}\right\rangle [1+\cos^2\theta]+
\left\langle\frac{(\triangle J_{\|})^2}
                 {\triangle t}\right\rangle \sin^2\theta \right )~~~,
\nonumber\\
\left\langle\frac{(\triangle J_{z})^2}
		 {\triangle t}\right\rangle &=&
\left\langle\frac{(\triangle J_{\|})^2}
		 {\triangle t}\right\rangle \cos^2\theta
+\left\langle\frac{(\triangle J_{\bot})^2}
		  {\triangle t}\right\rangle \sin^2\theta~~~,
\label{a.3}
\end{eqnarray}
where $i=x,y$ and all $\left\langle\frac{\triangle J_{i}\triangle J_{j}}
{\triangle t}\right\rangle$, $i\neq j$, vanish.

The ``precession averaging'' we perform here is  similar to the
``Larmor averaging'' used in Roberge et al. (1993). The difference is
that we allow for the increments of angular momentum along all axes,
while only increments along $z^b$-axis were accounted for 
in Roberge et al. (1993).

\subsection{Averaging over precession about magnetic field}

Due to the Barnett effect, a spinning grain develops magnetic moment 
anti-parallel to grain angular velocity (Dolginov \& Mytrophanov
1976, Purcell 1979). This magnetic moment interacts with the ambient
magnetic field and causes grain precession on the time-scale of several years.

Although the Larmor precession is much slower than the precession of
${\bf z}^b$ about $\bf J$, it 
is fast compared to the rate of gaseous damping and paramagnetic
alignment. Therefore the ``Larmor averaging'' is necessary. This averaging
is very similar to the one used above.

The Larmor precession is characterized by angle $\beta$. 
Let us introduce a system of reference $x_0y_0z_0$ with $z_0$-axis
along magnetic field and  $x_0$ and $y_0$ axes lying 
in the plane perpendicular to the field. 
It is easy to obtain quadratic coefficients
\begin{eqnarray}
\left\langle\frac{(\triangle J_{i})^2}{\triangle t}\right\rangle &=&
\frac{1}{2}
\left\langle\frac{(\triangle J_{\bot})^2}{\triangle t}\right\rangle 
\left(\frac{1}{2}
[1+\cos^2\theta][1+\cos^2\beta]+\sin^2\beta\sin^2\theta \right)
\nonumber \\
 &+&
\frac{1}{2}\left\langle\frac{(\triangle J_{\|})^2}
                    {\triangle t}\right\rangle \left(
\frac{1}{2}\sin^2\theta
[1+\cos^2\beta]+\cos^2\theta\sin^2\beta \right)~~~,
\nonumber\\
\left\langle\frac{(\triangle J_{z_0})^2}{\triangle t}\right\rangle &=&
\left\langle\frac{(\triangle J_{\|})^2}{\triangle t}\right\rangle 
\left(\cos^2\theta \cos^2\beta+\frac{1}{2}\sin^2\theta \sin^2\beta
\right) \nonumber\\
&+&
\left\langle\frac{(\triangle J_{\bot})^2}{\triangle t}\right\rangle 
\left(\sin^2\theta\cos^2\beta+\frac{1}{2}[1+\cos^2\theta]\sin^2\beta\right)~~~,
\label{b.3}
\end{eqnarray}
where $i=x_0,y_0$. The substitution of expressions for 
$\left\langle\frac{(\triangle J_{\bot})^2}{\triangle t}\right\rangle$,
$\left\langle\frac{(\triangle J_{\|})^2}{\triangle t}\right\rangle$ in Eq.~(\ref{b.3})
provides
\begin{eqnarray}
A &=& \frac{\sqrt{\pi}}{3}nmb_{m}^{4}v_{\it th}^{3}
\left(1+\frac{T_{\rm s}}{T_{g}}\right)~~~, \nonumber\\
\left\langle\frac{(\triangle J_{i})^2}{\triangle t}\right\rangle &=&
A \Gamma_{\bot}({\rm e}_m)
\left[\frac{1}{2}
[1+\cos^2\theta][1+\cos^2\beta]+\sin^2\beta\sin^2\theta \right]~~~,
\nonumber\\
&+&
A\Gamma_{\|}({\rm e}_m)
\left[\frac{1}{2}\sin^2\theta
[1+\cos^2\beta]+\cos^2\theta\sin^2\beta \right] 
\nonumber\\
\left\langle\frac{(\triangle J_{z_0})^2}{\triangle t}\right\rangle &=&
2A \Gamma_{\|}({\rm e}_m)
\left[\cos^2\theta \cos^2\beta+\frac{1}{2}\sin^2\theta \sin^2\beta
\right]
\nonumber\\
&+&
2A \Gamma_{\bot}({\rm e}_m)
\left[\sin^2\theta\cos^2\beta+\frac{1}{2}[1+\cos^2\theta]\sin^2\beta \right]
~~~,
\label{b.4}
\end{eqnarray}
where $i=x_0,y_0$. 
If $\theta=0$, it is easy to see that our coefficients coincide with
those in Roberge et al. (1993).

According to Eq.~(\ref{a.2.e1}), the damping is given by a vector
anti-parallel to $\bf J$. Therefore 
\begin{equation}
\left\langle \frac{\triangle J_{i}}{\triangle t}\right\rangle =
-\frac{J_i}{t_{\rm eff}}~~~,
\label{aaa11}
\end{equation}
where $i=x_0, y_0, z_0$.

The coefficients above can be used to solve the Fokker-Planck equation
numerically similarly to what was done in Roberge et al. (1993). 
Instead  we will use a perturbative approach
similar to  one introduced in Paper I. This approach was shown to
provide good accuracy when the Barnett relaxation
is complete and therefore we use it here.  In future we plan  to compare 
our analytical results with direct numerical simulations when
these simulations become available.

\section{Paramagnetic relaxation}

The diffusion coefficients above characterize gas-grain 
interactions. The effect of magnetic field with intensity $B$ 
on grains is imprinted through
the ``magnetic'' coefficients, their simplest form in $x_0 y_0 z_0$ reference 
frame is (see Jones \& Spitzer 1967) 
\begin{eqnarray}
\left\langle \frac{\bigtriangleup J_{j}}
		   {\bigtriangleup t}\right\rangle_{\rm mag}&=&
-\frac{J_{j}}{t_{\rm mag}}~~~,
\label{d27}\\
\left\langle \frac{\bigtriangleup J_{z_0}}
		  {\bigtriangleup t}\right\rangle_{\rm mag}&=&0~~~,
\label{d28}
\end{eqnarray}
where $j=x_0,y_0$, 
\begin{equation}
t_{\rm mag}=\frac{I_{z}^{b}}{\kappa VB^{2}}~~~,
\label{d29}
\end{equation}
with $V$ denoting grain volume and 
$\kappa\approx 2.5\cdot 10^{-12}T_{\rm s}^{-1}$~s used  for slow rotation
(Spitzer 1978). Similarly
\begin{eqnarray}
\left\langle\frac{(\bigtriangleup J_{j})^{2}}
		 {\triangle t}\right\rangle_{\rm mag}&=&
2k T_{\rm s}VB^{2}\kappa~~~,
\label{d30}\\
\left\langle\frac{(\bigtriangleup J_{z_0})^{2}}
		  {\triangle t}\right\rangle_{\rm mag}&=&0~~~,
\label{d300}
\end{eqnarray}
where $j=x_0,y_0$. Note, that Eqs~(\ref{d28}) and (\ref{d300}) reflect the 
fact that magnetic field does not slow down grains rotating about the field.

Only  the component of magnetic field perpendicular to $\bf J$ 
contributes to the Davis-Greenstein relaxation. The component 
parallel to $\bf J$ contributes to the relaxation in the grain frame of
reference.\footnote{This process can be called Jones \& Spitzer relaxation, 
as it was first mentioned in  Jones \& Spitzer (1967).}

\section{Perturbative approach}
\subsection{Small parameters}

It is well known that asymptotic results are usually 
attainable if there is a small parameter in a model. The difficulty in 
direct solving the Fokker-Planck equation comes from the fact that 
both $\left\langle\frac{\bigtriangleup J_{i}}{\triangle t}\right\rangle$ 
and  
$\left\langle\frac{(\bigtriangleup J_{i})^{2}}{\triangle t}\right\rangle$ 
depend on two variables, namely,  $\theta$ and $\beta$.

In general, $\Gamma_{\|}$ differs from $\Gamma_{\perp}$, but according to 
Paper~I, their ratio is close to unity and therefore 
\begin{equation}
\gamma=1-\frac{\Gamma_{\perp}}{\Gamma_{\|}}
\end{equation}
does not exceed 0.2. Using $\gamma$, it is possible to rewrite Eqs~(\ref{b.4})
so that terms containing $\gamma$ are clearly separated
\begin{eqnarray}
\left\langle \frac{(\bigtriangleup J_j)^{2}}
		  {\bigtriangleup t}\right\rangle &=&
2 A\Gamma_{\|}[1-\gamma\eta_2(\beta, \theta)]~~~,\nonumber\\
\left\langle \frac{(\bigtriangleup J_{z_0})^{2}}{\bigtriangleup t}
\right\rangle &=& 2A\Gamma_{\|}[1
-\gamma \eta_1(\beta, \theta)]~~~, \nonumber\\
\end{eqnarray}
where $j = x_0, y_0$ and 
\begin{eqnarray}
\eta_1(\beta, \theta)&=& 
\sin^2\theta+\sin^2\theta (\cos^2\beta-0.5\sin^2\beta)
\label{d36}~~~,\\
\eta_2(\beta, \theta) &=& \frac{1}{2}([1+\cos^{2}\theta]+
\sin^2\theta [\cos^2\beta-0.5\sin^2\beta])~~~.
\label{d34}
\end{eqnarray}
It is easy to see that both $\eta_1$ and  $\eta_2$ do not exceed unity for
all possible values $\theta$ and $\beta$. Therefore the problem of perturbative
treatment of these coefficients is very
similar to that used in Paper~I.

A new feature of our present study as compared with Paper I is the
dependence of the linear coefficient given by Eq.~(\ref{a.2.e1}) on $\theta$.
When $\theta\ll 1$, the variations of $\xi(\theta)$ (see Eq.~(\ref{a.2.e}) 
are small. The upper bound for $\theta$ can be found 
by comparing our predictions with direct numerical simulations
whenever such simulations become available. Therefore for
the time being, we study paramagnetic alignment assuming that
$\theta$ is our second small parameter. Fortunately, for many
cases of practical importance $\theta$ is small (LR).

\subsection{Asymptotic of the Barnett relaxation}

If $\theta\neq 0$,  $\bf J$ precesses in body coordinates. Purcell (1979)
was the first to realize that this must entail internal dissipation of
energy on relatively short time scales. According to Purcell (1979) the
most important mechanism of internal dissipation is the Barnett
dissipation\footnote{Our preliminary study shows that this is not
always true and Purcell (1979) underestimated the efficiency of
inelastic dissipation. This issue, however, does not change our
conclusions here and we will discuss inelastic dissipation elsewhere.}. This 
relaxation arises from alternating magnetization caused by precessing
Barnett moment (see previous section). For a thermally rotating 
grain of size $a=10^{-5}$ cm, the Barnett-equivalent
magnetic field is $\omega/\mu_r\approx 10^{-2} a_{-5}^{5/2}$~G,  
which is substantially stronger than interstellar magnetic field.\footnote{
In some  circumstances, e.g. in stellar atmospheres the external
magnetic field may become greater than the ``Barnett induced'' one.
In this case the relaxation is mainly caused by the component of 
external magnetic field parallel to $\bf J$. This, however,
may not alter our results here, provided that the relaxation
happens on the time-scale much less than the time of gaseous damping.}
Therefore
the rate of relaxation, which scales as $B^2$, is
nearly $10^{6}$ times faster than the paramagnetic alignment.

An elaborate study of the partial Barnett alignment was done in
LR for different ratios of Barnett  and gaseous damping times.
Here we consider the alignment when this ratio approaches zero. 
In this regime the value of $\theta$
is determined only by thermal fluctuations within grain material and for a
fixed angular momentum, is given by the Boltzmann distribution (LR)
\begin{equation}
f(\theta)={\rm const} \times 
\sin\theta \exp\left[-E_{\rm rot}(\theta)/kT_s\right]~~~,
\label{f_beta}
\end{equation}
where const is determined by normalization, and  grain rotational energy
is
\begin{equation}
E_{\rm rot}(\theta)=\frac{J^2}{2I_z}\left[1+(h-1)\sin^2\theta\right]~~~,
\label{e_beta}
\end{equation}
where $h=I_z/I_{\bot}$.

The mean value of $\cos\theta$ can be found as
\begin{equation}
\left\langle \cos\theta\right\rangle_J=
\frac{\int^{\pi}_{0}\cos^2\theta \sin\theta f(\theta)
{\rm d}\theta}{\int^{\pi}_{0}\sin\theta f(\theta){\rm d}\theta}~~~,
\label{ddis}
\end{equation}
where the subscript $J$ indicates that the value for is
calculated for an individual grain with a fixed angular momentum.

The value of $\bf J$ varies for different grains within an ensamble. 
 To find the mean value of $\cos\theta$ for the ensemble of grains, 
$\left\langle \cos\theta\right\rangle$, 
one has to average over the distribution
of $J$. However, numerical studies of Barnett alignment in LR95
have shown that with a high degree of accuracy 
one can find $\left\langle \cos\theta\right\rangle$ by substituting the mean
value of angular momentum corresponding to rotational temperature $T$
 in Eq.~(\ref{ddis}). This phenomenological
fact greatly simplifies our treatment and here we want
to provide a theoretical justification for it  by studying assymptotics
of the Barnett relaxation in the limit $\theta\ll 1$. 

Consider the evolution of $\bf J$ in  $xyz$ reference system. 
The contribution of the Barnett related coefficients
in this system is zero, as the Barnett relaxation
does not change the value of $J$. Therefore it is possible
to formulate the Fokker-Planck equation with the linear coefficient given by
Eq.~(\ref{a.2.e1}) and the quadratic coefficient given by the sum of
the quadratic coefficients (\ref{a.3}):
\begin{equation}
\left\langle\frac{(\triangle J)^2}{\triangle t}\right\rangle =3A\Gamma_{\|}
[1-\gamma/3 ]~~~.
\label{tuff}
\end{equation}

For small $\theta$, the $\theta$-dependence of the linear 
coefficient is weak (\ref{a.2.e1}). 
Thus we fix angle $\theta=\theta_i\ll 1$
and find 
\begin{equation}
F(J)={\rm const} \times\exp\left(-\frac{J^2}{3kT_m I_z \xi(\theta_i)
(1-\gamma/3)}\right)~~~,
\label{funct}
\end{equation}
where
\begin{equation}
T_m=0.5(T_m+T_g)~~~.
\end{equation}

A study of Eq.~(\ref{funct}) reveals only a weak dependence of
$J$ on $\theta$ and this corresponds to calculations in LR performed
for different distributions of $\theta$. This means that 
our perturbative solution describes essential physics for $\theta\ll 1$.

Further on in the paper we use the phenomenological result of LR and 
calculate the measure of internal alignment $Q_{\rm J}$ by substituting
the ensemble averaged value of $J$ in Eq.~(\ref{ddis}). In the
absence of paramagnetic relaxation, the Maxwellian mean value of 
angular momentum should be used (Landau \& Lifshitz 1980):
\begin{equation}
J^2=kT_m(2I_{\bot}+I_z)~~~.
\label{jjj}
\end{equation}
In the presence of paramagnetic dissipation, this value
can still be used as the zeroth  approximation  for $\theta_0$ 
to obtain the zeroth order approximation for $t_{\rm eff}$ and
 diffusion coefficients. 

In general, paramagnetic relaxation diminishes the value
of $J$. Paramagnetic alignment decreases the component of $\bf J$ 
perpendicular
to magnetic field and does not affect the component of $\bf J$ parallel to
the  field. Therefore it is reasonable to 
conjecture that the component of $\bf J$ parallel to magnetic
field stays Maxwellian and
\begin{equation} 
J^2=\frac{kT_m(2I_{\bot}+I_z)}{3\left\langle
\cos^2\beta\right\rangle}
\label{jjj1}
\end{equation}
 can be substituted instead of Eq.~(\ref{ddis})
as a mean value of $J^2$. Equations~(\ref{jjj1}), (\ref{f_beta}), 
(\ref{e_beta}) and (\ref{ddis}) enable one to obtain higher order
iterations of  the alignment measures.

\subsection{Iterations}

 We start with assuming $\gamma=0$ and $\theta=\theta_0$. Then
variables in Eq.~(\ref{d32}) can be separated. The stationary 
equation for the $z$ component is
\begin{equation}
\frac{1}{2}\frac{\partial^{2}}{\partial J_{z_0}^{2}}
\left(\left\langle \frac{(\bigtriangleup J_{z_0})^{2}}
		   {\bigtriangleup t}\right\rangle f_{z_0}\right)-
\frac{\partial}{\partial J_{z_0}}
\left(\left\langle \frac{\bigtriangleup J_{z_0}}
			{\bigtriangleup t}\right\rangle f_{z_0}\right)=0~~~,
\label{d38}
\end{equation}
which has the solution
\begin{equation}
\ln f_{z_0}=-\frac{J_{z_0}^{2}}{t_{\rm eff}
\left\langle \frac{(\bigtriangleup J_{z_0})^{2}}
		  {\bigtriangleup t} \right\rangle}
+{\rm const}_{1}~~~.
\end{equation}
In the case of $x_0$ and $y_0$ components, one has to account 
for paramagnetic relaxation and the solutions take the form
\begin{equation}
\ln f_{j}=
-\frac{J_{j}^{2}}
      {\left\langle (\bigtriangleup J_{j})^{2}\right\rangle+
       \left\langle (\bigtriangleup J_{j})^{2}\right\rangle_{\rm mag}}
\left(\frac{1}{t_{\rm eff}}+\frac{1}{t_{\rm mag}}\right)+{\rm const}_{2}~~~.
\end{equation}
To characterize the relative importance of magnetic torque, we define
\begin{equation}
\delta_{i}=\frac{\left\langle
\frac{ \bigtriangleup J_{j}}{\bigtriangleup t}\right\rangle_{\rm mag}}
{\left\langle\frac{\bigtriangleup J_{j}}{\bigtriangleup t}\right\rangle}=
\frac{t_{\it eff}}{t_{\rm mag}}=
\frac{3}{4\sqrt{\pi}}
\frac{\kappa VB^{2}}{nmv_{\it th}b_{m}^{4}\Gamma_{\|}}\xi(\theta_i)~~~.
\label{d41}
\end{equation}
As $\xi(\theta) < 1$ (see Fig.~3), Eq.~(\ref{d41}) shows that the incomplete
alignment increases randomization for oblate grains.

For $\gamma=0$, $\theta=\theta_0$  and $\delta_{i}=\delta_{0}$,  
the problem is similar to that discussed in 
Jones \& Spitzer (1967). The solution is also similar
\begin{equation}
\cos^{2}\theta_{0}=\frac{1}{1-\aleph_{0}^{2}}
\left[1-\frac{\aleph_{0}}{\sqrt{1-\aleph_{0}^{2}}}
{\rm arcsin}\sqrt{1-\aleph_{0}^{2}}\right]~~~,
\label{d42}
\end{equation}
where 
\begin{equation}
\aleph_{0}^{2}=\frac{1+\delta_{0}T_{\rm s}/T_{m}}
			 {1+\delta_{0}}~~~.
\label{aleph43}
\end{equation}

 For higher iterations modified diffusion coefficients are 
\begin{equation}
t_{\rm eff}\left\langle\frac{(\bigtriangleup J_{z_0})^{2}}
			    {\bigtriangleup t}\right\rangle=
2kT_mI_z\xi(\theta_i)(1-\gamma\eta_1(\beta_i,\theta_i))~~~,
\end{equation}
\begin{equation}
t_{\rm eff}\left(\left\langle
\frac{(\bigtriangleup J_{j})^{2}}{\bigtriangleup t}\right\rangle
+\left\langle\frac{(\bigtriangleup J_{j}^{2})}
		  {\bigtriangleup t}\right\rangle_{\rm mag}
\right)=2k I_{z_0}\xi(\theta_i)\left(T_m+\delta_{i}T_d \right)\left(1-
\gamma\frac{1+\eta_{2}(\beta_i, \theta_i)}{1+\delta_{i}T_d/T_m}\right)~~~,
\end{equation}
\begin{equation}
t_{\rm eff}\left\langle
\frac{(\bigtriangleup J_{j})^{2}}{\bigtriangleup t}\right\rangle_{\rm mag}=
2 kT_{\rm s}\delta_{1}I_{z}^{b}~~~,
\end{equation}
where $j=x_0,y_0$.

Thus, the distribution function for the $i$-th iteration is
\begin{equation}
f=f_{x_0}f_{y_0}f_{z_0}=
{\rm const}\cdot\exp\left\{\frac{J^2\left(1-\cos^2\beta\left[1-\aleph^2_i
\right]\right)}
{2k\xi(\theta_i)I_zT_{av}\left(1-\gamma W_i
\right)}\right\}~~~,
\label{distr}
\end{equation}
where
\begin{equation}
T_{\rm av}=\frac{T_{m}+T_{\rm s}\delta_{i-1}}{1+\delta_{i-1}}~~~,
\label{t_av}
\end{equation}
\begin{equation}
W_i=\frac{1+\eta_2(\beta_{i-1},\theta_{i-1})}{1+\delta_{i-1}T_d/T_m}.
\end{equation}
The solution for $\left\langle \cos^2\beta_i\right\rangle$ can 
be obtained by substituting
\begin{equation}
\aleph_i=\frac{T_{\rm av}
(1-\gamma W_i)}
{T_m[1-\gamma\eta_1(\beta_{i-1}, \theta_{i-1})]}
\label{aleph_i}
\end{equation} 
in Eq.~(\ref{d42}) for $\aleph_0$.

Using Eqs~(\ref{ddis}) and (\ref{d42}), we can find $i$-th iterations of 
$Q_{\rm X}$ and $Q_{\rm J}$ (see Eqs~(\ref{two}) and (\ref{three})). Then 
Eq.~(\ref{rail}) gives us
the Rayleigh reduction factor and this solves the problem.
It is possible to show that the corresponding formal series converge and
the error of the zero approximation does not exceed ten percent for
a wide range of ratios of magnetic to damping times and grain to gas
temperatures.
We stress ``formal'', however, as 
in our perturbative treatment we substitute particular values of $\theta_i$
and $\beta_i$ instead of the distributions of those angles. Therefore
a direct check of our results by testing against numerical simulations for 
a wide range of $\delta_i$ and $T_d/T_g$ is necessary. Although, 
the corresponding
numerical study is a challenging problem far more involved than the 
numerical simulations performed so far we hope that it can be
accomplished (Roberge \& Lazarian 1997, in preparation).
 
\subsection{Superparamagnetic grains}

An important case of grains that can be paramagnetically aligned is
``supergrains'' (grains with superparamagnetic or
superferromagnetic properties) suggested in Jones \& Spitzer (1967)
(see also Draine 1996).
Indeed, ordinary paramagnetic grains are only marginally
aligned in typical interstellar conditions\footnote{Here we speak about
alignment of grains with $a>10^{-5}$ cm, which are responsible
for the observed polarization (Kim \& Martin 1995). One may erroneously
assume that due to the dependence of $\delta_i$ on grain size (see Eq.~(\ref{d41}))
small grain should be efficiently aligned. For those grains, however,
one has to account for disorientation of grains 
through thermal emission and this decreases $\delta_i$.}
and therefore
cannot account for observed polarization. On the contrary,
``supergrains'' may be efficiently aligned and can account for
the peculiarities of the polarization curve (Mathis 1986, Martin see also
Goodman \& Whittet 1996).
Such grains rapidly dissipate their energy by interacting with
external magnetic field, and their ratio of gaseous to 
magnetic damping time is much greater than unity. 
Then $\delta_i$ given by Eq.~(\ref{d41}) is $\gg 1$ for any value
of $\theta$, and Eq.~(\ref{t_av}) provides
$T_{av}\approx T_d$, i.e. the dependence on $\theta$ cancels out.
As a result, for $\gamma=0$, Eq.~(\ref{aleph_i}) gives $\aleph\approx
T_d/T_m$, which coincides with the result obtained in Jones
\& Spitzer (1967) for spherical superparamagnetic grains.

For $\gamma\neq 0$, the iterations involve only one small parameter
and the problem is similar to that treated in Paper I. The 
parameter $\aleph_i$ can be written in the following way
\begin{equation}
\aleph_i^2\approx \frac{T_{\rm av}}
{T_m[1-\gamma \xi(\beta_i, \theta_i)]}\approx
 \frac{T_{\rm av}}{T_m}[1+\gamma \xi(\beta_i, \theta_i)]~~~,
\label{alll}
\end{equation}
where we disregarded  $\gamma W_i$ term on the account that 
 $W_i\ll 1$ when $\delta_i\gg 1$. For any $\theta$, $\xi(\theta)<1$ and 
formally there is no difference between Eq.~(\ref{alll}) and eq.~40
in Paper I. The latter expression for $\aleph$ resulted in the iteration
procedure that was proved accurate by numerical simulations\footnote{
According to private communication by W. Roberge some deviations were
observed in the limit of very small $\delta_i$. Such $\delta_i$ are of
minor practical importance, however.} (Roberge (1996),
DeGraff, Roberge \& Flaherty (1997)). Therefore we believe that
at least for superparamagnetic grains our analytical treatment is accurate.

\section{Discussion}

In comparing our present results with those
obtained in Paper I we want to define clearly the ranges of applicability  
of the corresponding formalisms.

Our results in Paper I are applicable to grains, which have rotational 
temperature much greater than the  temperature of 
grain material. In this limit the alignment of $\bf J$ in the grain 
frame of reference is almost perfect and the measure of alignment
can be obtained through the iteration procedure.

Apart from a trivial case of $T_g\gg T_d$ it is likely that the
treatment presented in Paper I is applicable if active sites 
of H$_{2}$ formation 
 cover the entire surface of grains and therefore H$_{2}$ formation 
 is essentially stochastic (Cugnon 1985). Such a chaotic formation of H$_{2}$
molecules
can make grains ``rotationally hot'' ($T_m\gg T_d$).  
Cosmic rays can also spin up grains in particular regions through
the process of momentum deposition described in
Purcell \& Spitzer (1971), though
really huge fluxes of cosmic rays are needed for the purpose.

As compared with Paper~I, the formalism presented in the present paper
is more versatile. Indeed, it allows to find
the measure of the Davis-Greenstein alignment for finite 
$T_m/T_g$ ratios. Unlike all earlier papers on the Davis-Greenstein
alignment the present one accounts for a recently discovered phenomenon
of incomplete internal relaxation. This enables one to find the
Rayleigh reduction factor for oblate grains with high accuracy.

Our results are especially important for superparamagnetic and 
superferromagnetic grains. Fortunately, for such grains the formal
treatment of the problem of $\bf J$ alignment in respect to magnetic field
is similar to that in Paper I. The
accuracy of the latter treatment was proved through numerical
simulations and this makes us optimistic about the applicability of
our results to ``supergrains'' for a wide range of grain and  gas
temperatures.

For ordinary paramagnetic grains, our analytical results are obtained
in the limit of small deviations of $\bf J$ from the axis of major
inertia. Although formally it is
possible to iterate for arbitrary deviations and obtain converging
series for the solution, we believe that more numerical studies 
are needed to test our results for finite $\theta$.

\acknowledgements

This work owes much to my fruitful discussions with Wayne Roberge and to 
his constructive criticism of my original idea. I am happy to acknowledge
very valuable imput by Roger Hildenbrand.
I am grateful to Bruce Draine and Lyman Spitzer for 
illuminating discussions and to  
Alyssa Goodman and Peter Martin for valuable comments. 
I acknowledge the support by
NASA grant NAG5 2858.

\clearpage
{\bf Figure Captions}

{\bf Fig.1} Euler angles are shown on this figure. The $z^{b}$ axis is directed
along the axis of major inertia, while  $z$-axis is directed along the
vector of angular momentum. Similar transformations involving Euler angles
are used to relate diffusion coefficients in the course of Larmor averaging.
In the latter case the equivalint of $z^b$-axis is directed along $\bf J$, 
while the equivalent of $z$-axis along magnetic field. Naturally, 
in the angle between these two axes is not $\theta$, but $\beta$.
Postcript file of this figure was given to us by Wayne Roberge.

{\bf Fig.2} The ratio of $t_{gas1}/t_{gas2}$ as a function of grain 
eccentricity. Zero eccentricity corresponds to spherical grains and
eccentricity $\rightarrow 1$ to flakes.

{\bf Fig.3} The ratio of $t_{\rm eff}/t_{gas1}$ as a function of $\theta$.
For sufficiently small $\theta$ this ratio is close to unity.


\begin{thebibliography}{}
\bibitem{} Cugnon, P. 1985, \aap, 152, 1
\bibitem{} Davis, L. \& Greenstein, J.L., 1951, \apj,  114, 206
\bibitem{} Davis, L. 1955, Vistas in Astronomy, ed. A. Beer, 1, 336
\bibitem[]{} Draine, B.T. 1996, in Polarimetry of the Interstellar Medium, 
eds Roberge W.G. and Whittet, D.C.B., p.16
\bibitem[]{} Draine, B.T., \& Weingartner J.C. 1997a ApJ, 470, 551.
\bibitem[]{} Draine, B.T., \& Weingartner J.C. 1997b ApJ, submitted
\bibitem{} Duley, W.W. 1978, \apj, 219, L129
\bibitem{} DeGraff, T.A., Roberge, W.E., \& Flaherty, J.E. 1997, \mnras,
(submitted)
\bibitem{} Dotson, J.L. 1997, \apj, (in press)
\bibitem{} Greenberg, J.M. 1968, in Nebulae and Interstellar Matter, 
Kuiper G.P. \& Middlehurst B.M. (eds.), Univ. of Chicago Press, vol. 7.
\bibitem{} Goodman, A.A., Jones, T.J., Lada, E.A., \&  Myers P.C. 1995, \apj, 
 448, 748
\bibitem{} Goodman, A.A., \& Whittet, D.C.B. 1995, \apj, 455, L181
\bibitem{} Hildebrand, R.H. 1988, {\it QJRAS},  29, 327
\bibitem{} Hildebrand, R.H., \& Dragovan, M. 1995, \apj, 450, 663 
\bibitem{} Jones, R.V., \& Spitzer, L.,Jr, 1967, \apj,  147, 943
\bibitem[]{} Kim, S.-H., \& Martin, P., G. 1995, ApJ, 444, 293 
\bibitem{} Landau, L.D., \& Lifshitz, E.M. 1976, Mechanics, Pergamon Press,
 Oxford
\bibitem{} Landau, L.D., \& Lifshitz, E.M. 1980, Statistical Physics (part I),
 Pergamon Press,
 Oxford, p. 86
\bibitem{} Lazarian, A.  1994, \mnras,  268, 713
\bibitem{} Lazarian, A.  1995a, \apj, 453, 229 (Paper I)
\bibitem{} Lazarian, A.  1995b, \apj, 451, 660
\bibitem{} Lazarian, A., \& Draine B.T. 1997, \apj (submitted)
\bibitem{} Lazarian, A. \& Roberge W.G. 1997, \apj accepted
\bibitem{} Lee, H.M., \& Draine, B.T. 1985, \apj, 290, 211
\bibitem{} Lifshitz, E.M., \& Pitaevskii L.P. 1981, Physical Kinetics,
 Pergamon Press, Oxford
\bibitem{} Martin, P.G. 1995, \apj, 445, L63 
\bibitem{} Mathis, J.S. 1986, \apj,  308, 281.
\bibitem{} Purcell, E.M. 1969, On the Alignment of Interstellar Dust, 
 Physica,  41, 100.
\bibitem{} Purcell, E.M. 1979, \apj, 231, 404.
\bibitem{} Purcell, E.M., \& Spitzer, L., Jr  1971, \apj,  167, 31.
\bibitem{} Reichl, L.E. 1980, Modern Course in Statistical Physics, 
Edward Arnold (Publishers) Ltd., p~168.
\bibitem{} Roberge, W.G. 1996 in Polarimetry of the Interstellar 
Medium, eds Roberge W.G. and Whittet, D.C.B. p. 401
\bibitem{} Roberge, W.G., DeGraff, T.A., \& Flaherty, J.E., 1993, \apj, 
 418, 287.
\bibitem{} Spitzer, L., Jr  1978, Physical Processes in the Interstellar
Medium, Wiley-Interscience Publ., New York.
\bibitem{} Whittet, D.C.B. 1992, Dust in the Galactic Environment, Bristol
Inst. Physics
\end{thebibliography}
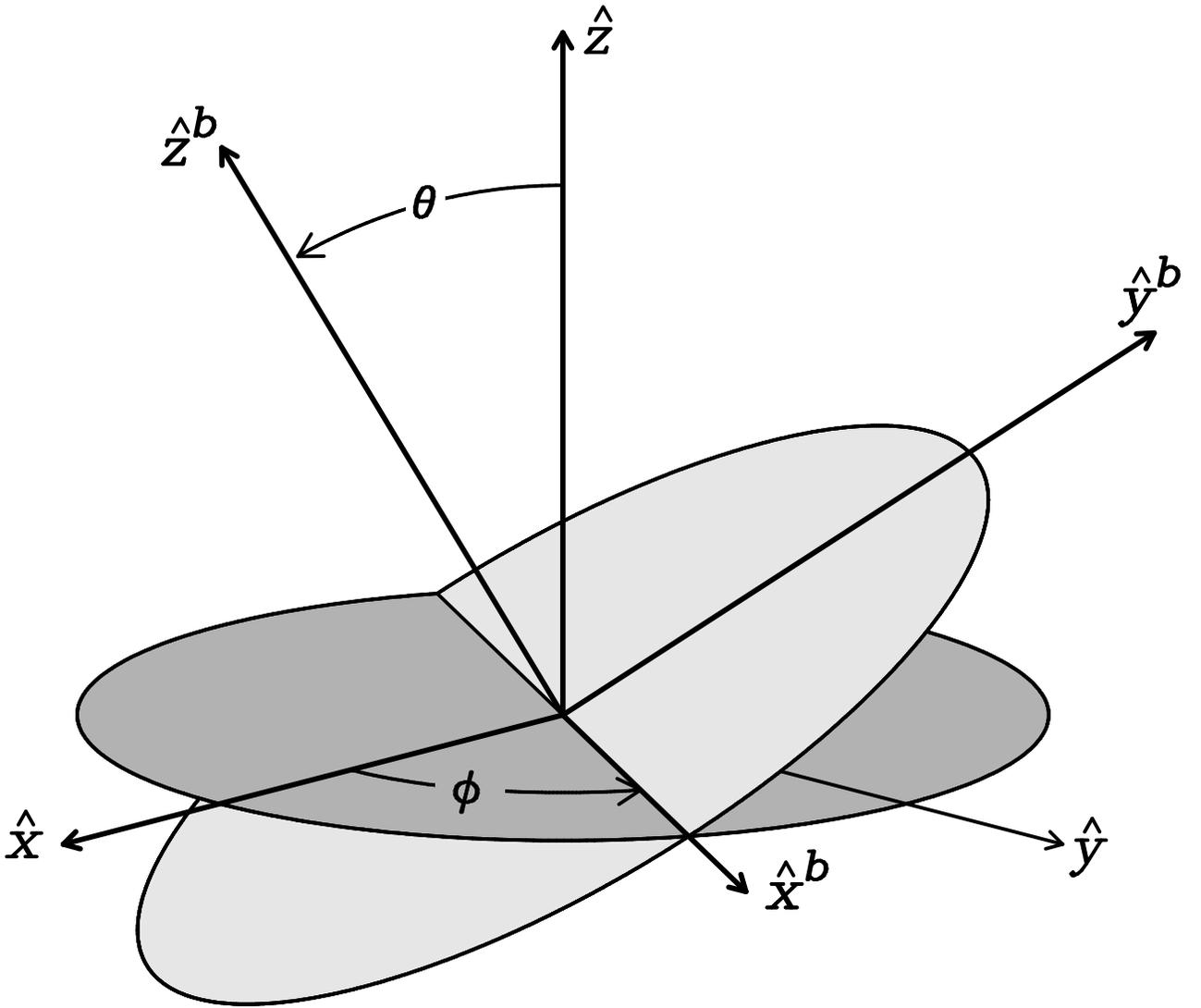
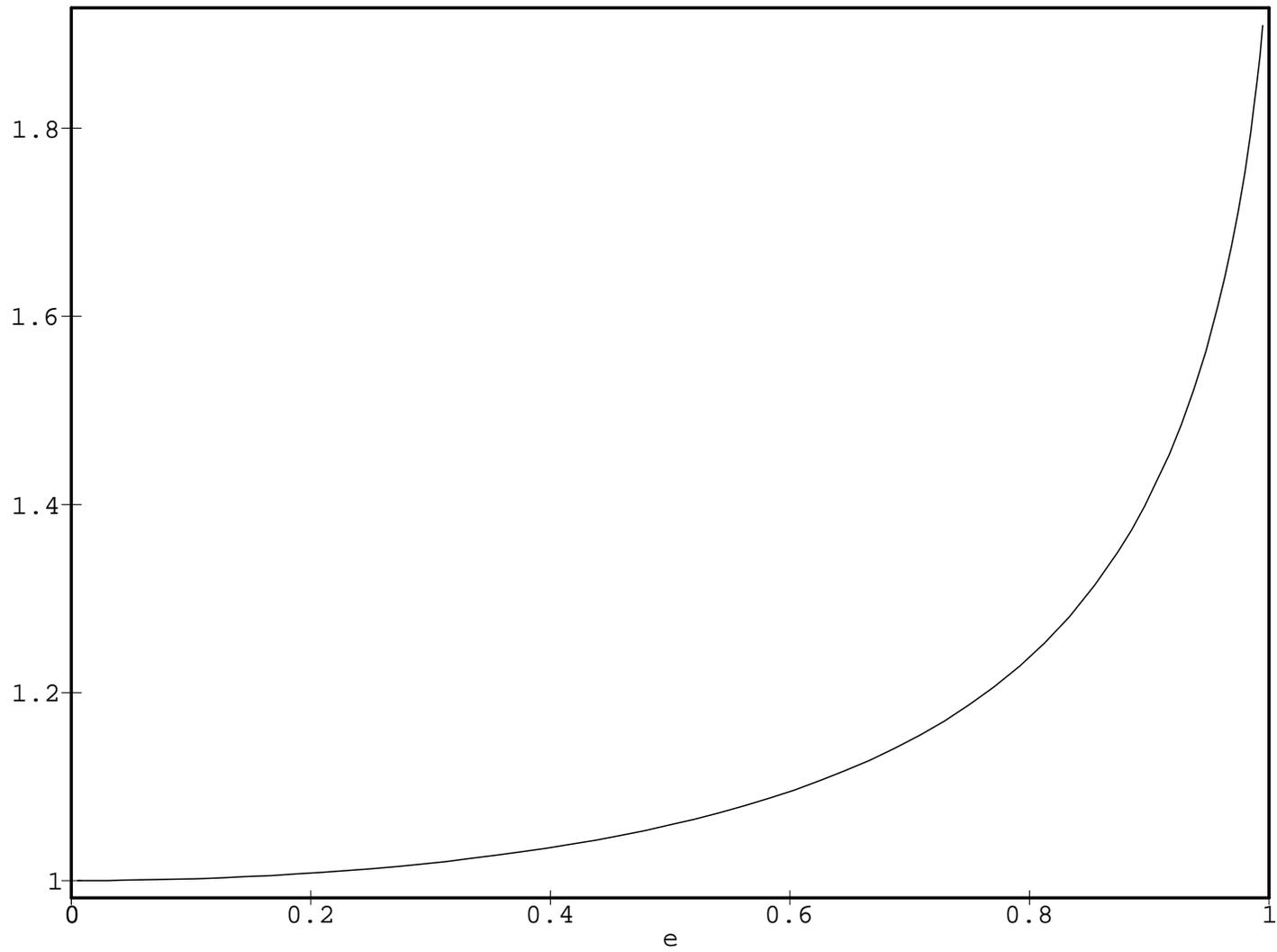
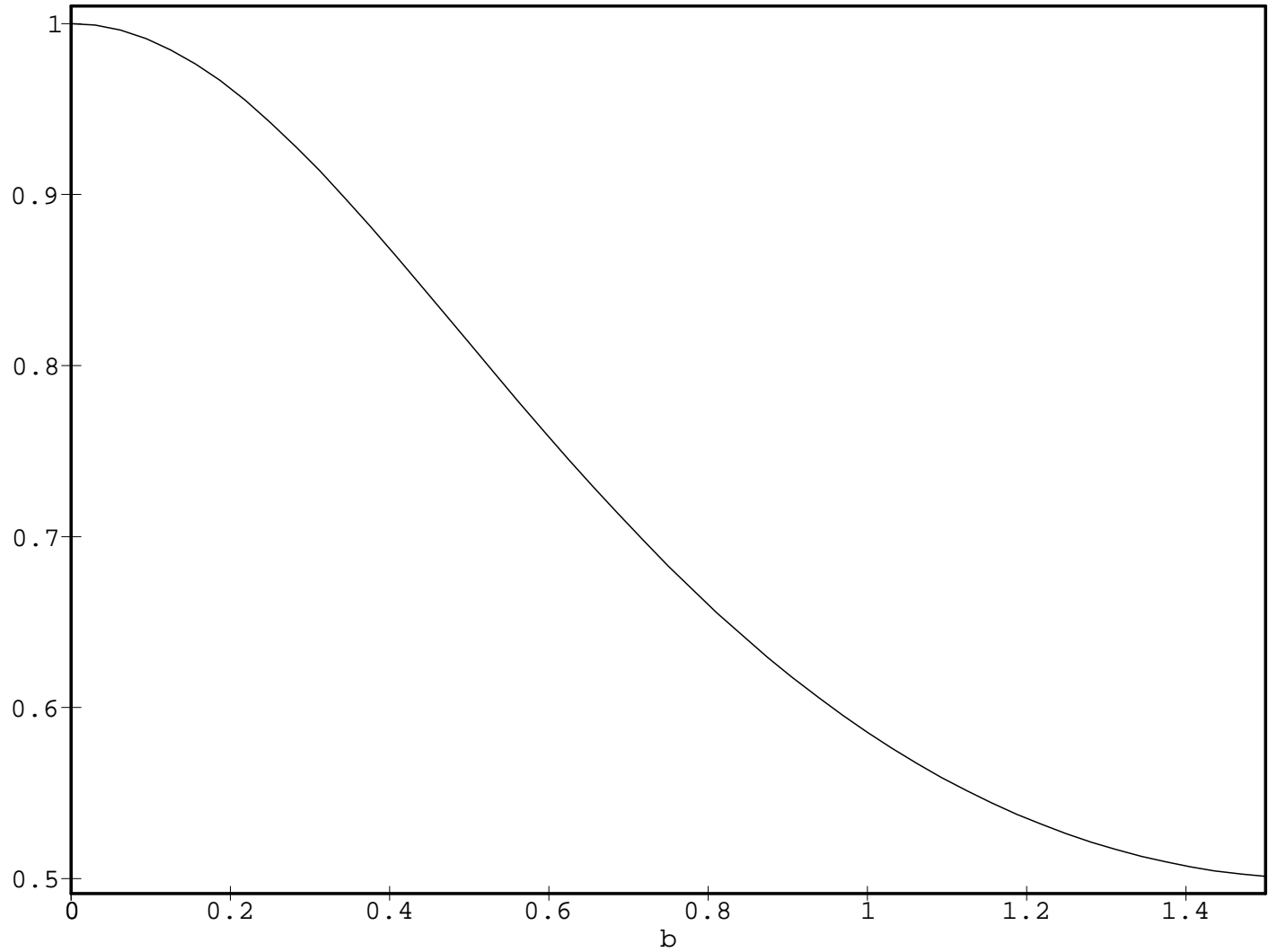
\end{document}